\documentclass[12pt]{article}%
\usepackage{amsmath}
\usepackage{amsfonts}
\usepackage{amssymb}
\usepackage{graphicx}%
\usepackage{setspace}
\usepackage{lineno}
\usepackage{geometry}
\usepackage[all]{nowidow}
\usepackage{array}
\providecommand{\U}[1]{\protect\rule{.1in}{.1in}}
\newtheorem{theorem}{Theorem}

\newtheorem{algorithm}[theorem]{Algorithm}

\newtheorem{problem}[theorem]{Problem}

\newenvironment{proof}[1][Proof]{\noindent\textbf{#1.} }{\ \rule{0.5em}{0.5em}}
\begin{document}
\title{\textbf{Sparsity Constrained Split Feasibility for Dose-Volume Constraints in
Inverse Planning of Intensity-Modulated Photon or Proton Therapy}}
\author{Scott Penfold$^{1,2}\thanks{\dag The contributions of the first two authors to
this work are of equal shares.}$, Rafa\l \ Zalas$^3\dag$, Margherita
Casiraghi$^4$
\and Mark Brooke$^2$, Yair Censor$^{5}$, Reinhard Schulte$^{6}$\medskip\\$^{1}$Department of Medical Physics, Royal Adelaide Hospital\\Adelaide, SA 5000, Australia \medskip\\ $^{2}$Department of Physics, \\University of Adelaide, Adelaide, SA 5005, Australia\medskip\\$^{3}$Department of Mathematics, The Technion, Technion City,\\Haifa 32000, Israel 
\medskip\\$^{4}$Paul Scherrer Institute, Center for Proton Therapy (CPT)\\Switzerland 
\medskip\\$^{5}$Department of Mathematics, University of Haifa\\Mt.\ Carmel, Haifa 3498838, Israel 
\medskip\\$^{6}$Department of Basic Sciences, School of Medicine \\Loma Linda University, Loma Linda, CA 92354, USA.\medskip\\
(scott.penfold@sa.gov.au)
}

\date{July 10, 2016; Revised January 10, 2017}
\maketitle
\newpage

\doublespacing
\begin{abstract}
\noindent A split feasibility formulation for the inverse problem of intensity-modulated radiation therapy (IMRT) treatment planning with dose-volume constraints (DVCs) included in the planning algorithm is presented. It involves a new type of sparsity constraint that enables the inclusion of a percentage-violation constraint in the model problem and its handling by continuous (as opposed to integer) methods. We propose an iterative algorithmic framework for solving such a problem by applying the feasibility-seeking $CQ$-algorithm of Byrne combined with the automatic relaxation method (ARM) that uses cyclic projections. Detailed implementation instructions are furnished. Functionality of the algorithm was demonstrated through the creation of an intensity-modulated proton therapy plan for a simple 2D C-shaped geometry and also for a realistic base-of-skull chordoma treatment site. Monte Carlo simulations of proton pencil beams of varying energy were conducted to obtain dose distributions for the 2D test case. A research release of the $Pinnacle^3$ proton treatment planning system was used to extract pencil beam doses for a clinical base-of-skull chordoma case. In both cases the beamlet doses were calculated to satisfy dose-volume constraints according to our new algorithm. Examination of the dose-volume histograms following inverse planning with our algorithm demonstrated that it performed as intended. The application of our proposed algorithm to dose-volume constraint inverse planning was successfully demonstrated. Comparison with optimized dose distributions from the research release of the $Pinnacle^3$ treatment planning system showed the algorithm could achieve equivalent or superior results.
\end{abstract}


\newpage

\section{Introduction \label{sect:introduction}}

Intensity-modulated radiation therapy (IMRT) with photons or intensity-modulated proton therapy (IMPT) are rapidly evolving techniques for planning and delivering radiation therapy to solid tumors. For many tumor sites, IMRT with photons has superseded standard radiation therapy (RT) techniques and is becoming the new standard in RT delivery \cite{Coles200516,Mohan2000619,Narayana2006892,Nutting2000649,Sura200817}. At existing proton centers, IMPT in combination with active pencil beam scanning is increasingly being used, replacing older passively scattered and collimated proton therapy techniques as a means for more accurately delivering high doses to the target volume and sparing of organs at risk (OARs) as indicated by dosimetric studies \cite{Lomax2004147,Kooy201588,vanDijk201611,Welsh20111336,Zhang2010357}. 

Instead of using a single upper dose bound for OARs and single lower dose bound for the target volumes, it has become a common practice in clinical trials and off-trial photon IMRT treatments to specify more than one dose-volume constraint (DVC), allowing a certain percentage of volume to violate to a certain extent a given bound. This additional DVC, which could be single or multiple, rely on accumulated clinical experience with conformal RT techniques. For example, Gulliford et al. \cite{Gulliford2010747} performed a detailed dose-volume analysis of the incidence of clinically relevant late rectal toxicities in patients treated with high-dose photon IMRT for prostate cancer and found that the incidence of moderate-to-severe rectal toxicity for any of six late-toxicity endpoints decreased incrementally for patients whose treatment plans met increasing numbers of DVCs from the set of $V_{30}\leq80\%$, $V_{40}\leq65\%$, $V_{50}\leq55\%$, $V_{60}\leq40\%$, $V_{65}\leq30\%$, $V_{70}\leq15\%$, and $V_{75}\leq3\%$. Here, $V_X \leq Y$ corresponds to a dose-volume constraint that $Y\%$ of the volume cannot receive more than $X$ Gy. These and similar DVCs for OARs have found their way into clinical trial protocols and practice guidelines over the years, see, e.g., \cite{RTOG1308}.

Most modern inverse planning algorithms attempt to incorporate DVCs by defining sub-volumes with different dose objectives applied to each sub-volume. The multiple objectives are then combined into a single cost function to be minimized. Minimization in RT inverse planning with DVCs has been performed with a number of different approaches. Spirou and Chui \cite[Section F]{sc98} used gradient descent to seek a vector of ray intensities that minimized a cost function representing the sum of all dose constraints violations. However, incorporating DVCs directly into the cost function of the minimization process often renders the objective function non-convex and non-differentiable. This has the disadvantage of potentially resulting in local minima and thereby sub-optimal treatment plans. Cho \textit{et al.} \cite{clmosp1998} used a similar concept but applied simulated annealing for minimization. Simulated annealing is less susceptible to non-convexity and non-differentiability but is less computationally efficient than gradient descent. Romeijn \textit{et al.} \cite{romeijn2003,romeijn2006} adopted a linear programming approach to handle what they called partial-volume constraints. However to make the problem tractable for computation, they replaced the familiar concept of DVCs by a closely related, but not identical notion of conditional value-at-risk (C-VaR). Zhang and Merritt \cite{zm2008} proposed a new least-squares model to handle DVCs while retaining differentiability at the expense of having to deal with a nested double minimization problem. Therefore, an inverse planning algorithm for DVCs that is computationally efficient, robust to non-convexity and non-differentiability yet without simplifying the problem statement has yet to be developed.


In the current work, feasibility-seeking methods, as opposed to minimization algorithms, are applied to RT inverse planning with DVCs. Within the proposed feasibility-seeking approach issues of convexity and differentiability of the cost function do not arise at all because no cost function is used. While the DVCs do require a constraint that is not convex (the sparsity-norm constraint set), we are able to incorporate it into the projection method that we use to solve the feasibility-seeking problem. This is possible because we have devised a way to calculate the projection onto this set in spite of it being non-convex.



Another general advantage of the feasibility-seeking approach has to do with the availability of a class of highly efficacious feasibility-seeking \textit{projection methods}. These methods refer to iterative algorithms that use projections onto sets while relying on the general principle that when a family of, usually closed and convex, constraints sets is present, then projections onto the individual sets are easier to perform than projections onto other sets (intersections, image sets under some transformation, etc.) that are derived from the individual sets. Furthermore, projection methods may have algorithmic structures that are particularly suited for parallel computing, such as block-iterative projections (BIP) or string-averaging projections (SAP). They also demonstrate desirable convergence properties and good initial behavior patterns. See, for example, the 1996 review \cite{bb96}, the recent annotated bibliography of books and reviews \cite{cc15} and its references, and \cite{coap}.





We recently showed that IMPT inverse planning is possible with a fully-discretized, feasibility-seeking approach by iteratively projecting solution vectors in the beam intensity vector space onto half-spaces representing dose constraints in target and OAR volumes \cite{Penfold2015}. In our preliminary work, we demonstrated that with these iterative projection algorithms, feasible solutions meeting the planning objectives can be found that meet target and normal tissues dose bounds, in particular, if the constraints are not too challenging and/or the treatment modality is very conformal (e.g., by using protons).

In this paper, we use the fully-discretized feasibility-seeking approach applicable to either photon IMRT or IMPT inverse planning which leads to a mathematical feasibility problem. The upper and lower bounds on the doses to the various structures define the linear inequality constraints of the feasibility problem, which is solved by feasibility-seeking projection methods without attempting to minimize any cost function. Within this setup, we propose and investigate a novel method for allowing the feasibility-seeking inverse planning algorithm to automatically account for DVCs.

In the next section, we rigorously define the notion of percentage-violation constraint (PVC), which does not seem to have been used in the mathematical optimization community until now. A PVC injects integers into the problem which makes it difficult to solve. To circumvent this difficulty, we reformulate the PVC with the aid of a sparsity norm that counts the number of non-zero entries in a vector. This enables us to replace the original feasibility problem with PVC by another feasibility problem that includes non-convex constraints for the sparsity norm. For the resulting feasibility problem with this non-convex sparsity norm induced constraint we develop a new iterative projection algorithm which is a combination of the $CQ$-algorithm \cite{Byrne2002} and the automatic relaxation method (ARM) \cite{arm}.

\section{Methods}
\subsection{Linear feasibility with percentage-violation constraints\label{sect:LFP-PVC}}

Given $p$ closed convex subsets $Q_{1},Q_{2},\cdots,Q_{p}\subseteq R^{n}$ of
the $n$-dimensional Euclidean space $R^{n}$, expressed as level sets%
\begin{equation}
Q_{j}=\left\{  x\in R^{n}\mid f_{j}(x)\leq v_{j}\right\}  ,\text{ for all
}j\in J:=\{1,2,\ldots,p\}, \label{eq:set}%
\end{equation}
where $f_{j}:R^{n}\rightarrow R$ are convex functions and $v_{j}$ are some
given real numbers, the convex feasibility problem (CFP) is to find a point
$x^{\ast}\in Q:=\cap_{j\in J}Q_{j}.$ If $Q=\emptyset$ where $\emptyset$ is the empty set then the CFP is said to
be inconsistent.

Denoting the inner product of two vectors in $R^{n}$ by $\left\langle a,b\right\rangle:=\sum_{i=1}^{n}{a_i b_i}$, we consider the following linear feasibility problem (LFP) with
percentage-violation constraint (PVC).

\begin{problem}
\label{prob:1}\textbf{Linear Feasibility with Percentage-Violation
Constraint (PVC).} Given a CFP as in (\ref{eq:set}) with $f_{j}(x)=$ $\left\langle
a^{j},x\right\rangle $ and two real numbers $0\leq\alpha\leq1$ and
$0<\beta<1,$ find a vector $x^{\ast}$ that solves the system%
\begin{equation}
\left\langle a^{j},x\right\rangle \leq v_{j},\text{ for all }j\in J
\label{eq:original-ineq}%
\end{equation}
subject to the additional PVC constraint that:%
\begin{align}
&  \mathbf{In\ up\ to\ a}\text{ }\mathbf{fraction\ \alpha}\text{
}\mathbf{(i.e.,}\text{ }\mathbf{100\alpha\%)}\text{ }\mathbf{of\ the\ total}%
\text{ }\mathbf{number}\text{ }\mathbf{of}\text{ }\mathbf{inequalities}%
\nonumber\\
&  \mathbf{in}\text{ (\ref{eq:original-ineq}) }\mathbf{the\ right}%
\text{-}\mathbf{hand\ side\ bounds}\text{ }v_{j}%
\mathbf{\ may\ be\ potentially\ violated}\nonumber\\
&  \mathbf{by\ up\ to\ a}\text{ }\mathbf{fraction\ \beta}\text{ }%
\mathbf{(i.e.,}\text{ }\mathbf{100\beta\%)}\text{ }\mathbf{of}\text{
}\mathbf{their}\text{ }\mathbf{values.} \label{eq;alpha-beta}%
\end{align}

\end{problem}

A PVC is an integer constraint by its nature. It changes the CFP to which it
is attached from being a continuous feasibility problem into becoming a mixed
integer feasibility problem. In the field of intensity-modulated radiation therapy (IMRT) treatment
planning dose-volume constraints (DVCs) are traditionally used to evaluate
treatment plans. DVCs are percentage-violation constraints but without
properly incorporating them into the algorithm itself it is not possible to a
priori guarantee that a solution will indeed obey them.

In this paper we propose a novel way to incorporate PVCs via the notion of a sparsity
norm and derive a tractable model and algorithmic approach, along with
detailed implementation instructions for using it, to solve DVCs feasibility
problems for inverse planning in IMRT.

\subsection{IMRT problem statement \label{sect:prob-statement}}

We consider the following linear interval feasibility problem (LIFP) which is
the basic model for the inverse problem in the fully-discretized approach to
IMRT treatment planning \cite{CENSOR1988,YC08,censor2003}:

\begin{problem}
\label{prob:basic-imrt}\textbf{Linear Interval Feasibility: the basic model
for the inverse problem in the fully-discretized approach to IMRT treatment
planning}. Find $x^{\ast}\in R^{n}$ for which the following hold:%
\begin{gather}
0\leq A_{1}x\leq b^{1},\label{eq:1}\\
b^{3}\geq A_{2}x\geq b^{2},\label{eq:2}\\
0\leq A_{3}x\leq b^{4},\label{eq:3}\\
x\geq0, \label{eq:4}%
\end{gather}
where $A_{1}\in R_{+}^{m_{1}\times n}$, $A_{2}\in R_{+}^{m_{2}\times n}$,
$A_{3}\in R_{+}^{m_{3}\times n}$ are given matrices, $b^{1}\in R_{+}^{m_{1}}$,
$b^{2},b^{3}\in R_{+}^{m_{2}}$, $b^{4}\in R_{+}^{m_{3}}$ are given vectors.
(The subscript + denotes the nonnegative orthant.)
\end{problem}

In IMRT the row inequalities of (\ref{eq:1}) represent voxels of an \textit{organ at
risk} (OAR) whose permitted absorbed doses should not exceed $b_{t}^{1}$ for
each voxel $t$ in this structure. The row inequalities of (\ref{eq:3}) represent voxels of another OAR whose
permitted absorbed doses should not exceed $b_{t}^{4}$ for each voxel $t$ in
this structure. The row inequalities of (\ref{eq:2}) represent voxels of a \textit{planning
target volume} (PTV) whose permitted absorbed doses should be above
$b_{t}^{2},$ but should not exceed $b_{t}^{3},$ for each voxel $t$ in this
structure.

Our tool to \textquotedblleft translate\textquotedblright\ the integer
constraint (\ref{eq;alpha-beta}) into a \textquotedblleft
continuous\textquotedblright\ one is the notion of sparsity norm, called
elsewhere the zero-norm, of a vector $x\in R^{n}$ which counts the number of
nonzero entries of $x,$ that is,%
\begin{equation}
\Vert x\Vert_{0}:=\left\vert \{x_{i}\mid x_{i}\neq0\}\right\vert ,
\end{equation}
where $\mid\cdot\mid$ denotes here the cardinality, i.e., the number of elements
of a set. This notion has been recently used for various purposes in
compressed sensing, machine learning and more. The \textquotedblleft lower $+$%
\ operation\textquotedblright\ on a vector $x\in R^{n}$ means that, for all
$i=1,2,\ldots,n,$%
\begin{equation}
(x_{+})_{i}:=\max(0,x_{i})=\left\{
\begin{array}
[c]{l}%
x_{i},\text{ if }x_{i}>0,\\
0,\text{ if }x_{i}\leq0.
\end{array}
\right.  .
\end{equation}
Obviously, $x_{+}$ is always a component-wise nonnegative vector. Hence,
$\Vert x_{+}\Vert_{0}$ counts the number of positive entries of $x$ and is
defined by%
\begin{equation}
\Vert x_{+}\Vert_{0}:=|\{x_{i}\mid x_{i}>0\}|.
\end{equation}

To incorporate a DVC related to (\ref{eq:1}) into the LIFP of Problem
\ref{prob:basic-imrt} we formulate another feasibility problem as follows.

\begin{problem}
\label{prob:imrt+dvc}\textbf{Linear Interval Feasibility with DVC for the
inverse problem in the fully-discretized approach to IMRT treatment planning}.
Find $x^{\ast}\in R^{n}$ for which%
\begin{gather}
0\leq A_{1}x\leq(1+\beta)b^{1},\label{eq:dvc1}\\
b^{3}\geq A_{2}x\geq b^{2},\label{eq:dvc2}\\
0\leq A_{3}x\leq b^{4},\label{eq:dvc3}\\
x\geq0,\label{eq:dvc4}\\
\Vert(A_{1}x-b^{1})_{+}\Vert_{0}\leq\alpha m_{1}, \label{eq:dvc5}%
\end{gather}
where $A_{1}$, $A_{2}$, $A_{3}$, $b^{1}$, $b^{2}$, $b^{3}$ and $b^{4}$ are as
in (\ref{eq:1})--(\ref{eq:4}), and $\beta>0$ and $\alpha\in\lbrack0,1]$ are
given real numbers.
\end{problem}

In this problem (\ref{eq:dvc1}) allows the doses to voxels of this
structure to \textquotedblleft overflow\textquotedblright\ by $\beta.$
(\ref{eq:dvc3}) represents an OAR to which we do not attach a DVC for now.
(\ref{eq:dvc2}) represents a PTV to which we do not attach a DVC for now.
(\ref{eq:dvc4}) are the nonnegativity constraints on the solution vector of intensities.

The novelty of the model lies in (\ref{eq:dvc5}). It says that since we demanded
originally $A_{1}x\leq b^{1}$ in (\ref{eq:1}) we must look at the
\textquotedblleft plussed difference vector\textquotedblright\ $(A_{1}%
x-b^{1})_{+}$. It is nonnegative and has a nonzero component exactly and only
in components that belong to row inequalities in (\ref{eq:dvc1}) for which
(\ref{eq:1}) is violated.

The zero-norm of $(A_{1}x-b^{1})_{+}$ is thus equal to the number of those
violations and (\ref{eq:dvc5}) restricts this number to be not greater than
$\alpha m_{1}$ where $m_{1}$ is the total number of row inequalities (i.e.,
voxels) in the OAR described by (\ref{eq:dvc1}). Thus, (\ref{eq:dvc5})
guarantees that the number of violations up to $\beta$ in (\ref{eq:dvc1})
remains at bay under the number $\alpha m_{1}.$ In the following we
propose to use an efficient iterative projections method to solve Problem
\ref{prob:imrt+dvc}.

\subsection{Projection methods for feasibility-seeking\label{sect:proj}}

Projections onto sets are used in a wide variety of methods in optimization
theory but here \textit{projection methods} refer to iterative algorithms that use
projections onto sets while relying on the general principle that when a
family of, usually closed and convex, sets is present, then projections onto
the given individual sets are easier to perform than projections onto other
sets (intersections, image sets under some transformation, etc.) that are
derived from the given family of individual sets.

Projection methods may have different algorithmic structures, such as
block-iterative projections (BIP) or string-averaging projections (SAP) of
which some are particularly suitable for parallel computing, and they
demonstrate nice convergence properties and/or good initial convergence
patterns. This class of algorithms has witnessed great progress in recent
years and its member algorithms have been applied with success to many
scientific, technological and mathematical problems. See, e.g., the 1996
review \cite{bb96}, the recent annotated bibliography of books and reviews
\cite{cc15} and its references, the excellent book \cite{CEG12}, or
\cite{coap}.

For the LIFP of Problem \ref{prob:imrt+dvc} one can use any of a variety of
projection methods to handle linear inequality constraints. The most famous of
those might be the Agmon-Motzkin-Schoenberg (AMS) cyclic feasibility-seeking
algorithm \cite{agmon,mot-schoen54}. In this paper we adopt a projection
method of a particular nature, namely, the automatic relaxation method (ARM)
for solving interval linear inequalities of \cite[Algorithm 1]{arm}.

ARM has two advantages over other projection methods applicable to this
problem: (i) it handles in each iteration an interval constraint and does not
need to handle the right-hand side and left-hand side inequalities of an
interval separately, (ii) additionally, it automatically implements a
relaxation strategy for the projections which takes into account how far from
the hyperslab, defined by an interval constraint, is the point that needs to
be projected on it and automatically and continuously adjusts the relaxation
parameter for the projection accordingly. The ARM generalizes the algebraic
reconstruction technique ART3 \cite{herman-ART3} and is further discussed
in Subsection 5.10 of Censor and Zenios \cite{CZ97}.

\subsection{Algorithmic approach\label{sect:alg-approach}}

First we observe that Problem \ref{prob:imrt+dvc} is a split feasibility
problem. Split feasibility
problems were introduced first in \cite{ce94} and further studied in
\cite{mssfp, cgr} and many other publications. The constraints (\ref{eq:dvc1})--(\ref{eq:dvc3}) can be collectively
described by $c\leq Ax\leq b$, where $A$ is an $(m_{1}+m_{2}+m_{3})\times n$
matrix composed from blocks%
\begin{equation}
A:=\left(
\begin{tabular}
[c]{l}%
$A_{1}$\\
$A_{2}$\\
$A_{3}$%
\end{tabular}
\ \right)  , \label{eq:A}%
\end{equation}
$b$ is an $(m_{1}+m_{2}+m_{3})$ vector given by%
\begin{equation}
b:=\left(
\begin{tabular}
[c]{l}%
$(1+\beta)b^{1}$\\
$b^{3}$\\
$b^{4}$%
\end{tabular}
\ \right)  , \label{eq:b}%
\end{equation}
and $c$ is an $(m_{1}+m_{2}+m_{3})$ vector given by%
\begin{equation}
c:=\left(
\begin{tabular}
[c]{l}%
$0$\\
$b^{2}$\\
$0$%
\end{tabular}
\ \right)  , \label{eq:c}%
\end{equation}
and they, along with (\ref{eq:dvc4}) all reside\ in the space $R^{n}$ of
intensity vectors $x.$ On the other hand, the sparsity constraint
(\ref{eq:dvc5}) takes place in the space $R^{m_{1}}$ where the vectors of
doses in the OAR (\ref{eq:1}) are, namely, the vector\ $b^{1}$ and the vectors
$y=A_{1}x.$ Therefore, we must use not plain feasibility-seeking methods but
feasibility-seeking methods for split feasibility problems.

In the space $R^{n}$ of intensity vectors we define the set%
\begin{equation}
C:=\{x\in R^{n}\mid c\leq Ax\leq b\}\cap R_{+}^{n} \label{eq:C}%
\end{equation}
where $A$, $b$ and $c$ are as in (\ref{eq:A}), (\ref{eq:b}) and (\ref{eq:c}),
respectively, and $R_{+}^{n}$ is the nonnegative orthant of $R^{n}.$ In
$R^{m_{1}}$, the space of dose vectors of the OAR structure represented by
(\ref{eq:1}), we define the set%
\begin{equation}
Q:=\{y\in R^{m_{1}}\mid\Vert(y-b^{1})_{+}\Vert_{0}\leq\alpha m_{1}\}
\label{eq:Q}%
\end{equation}
with $b^{1}$ and $\alpha m_{1}$ as in (\ref{eq:dvc1}). If a point $y=A_{1}x$
is in $Q$ then it is guaranteed to fulfil (\ref{eq:dvc5}). So, our split
feasibility problem is to find a point $x^{\ast}\in C$ such that $A_{1}%
x^{\ast}\in Q,$ precisely describing Problem \ref{prob:imrt+dvc} above.

Common feasibility or split feasibility problems deal with convex sets
but here we observe that $Q$ is not a convex set. However, we show below how
to project onto it orthogonally, thus enabling to use a
feasibility-seeking projection method for our Problem \ref{prob:imrt+dvc}.

To solve the split feasibility formulation of Problem \ref{prob:imrt+dvc} we propose to use the $CQ$-algorithm \cite{Byrne2002} for the sets $C$ and $Q$
given by (\ref{eq:C}) and (\ref{eq:Q}), respectively. It has the advantage
that it does not require to calculate the inverse $A_{1}^{-1}$ of $A_{1}$ in
order to \textquotedblleft go back\textquotedblright\ from $R^{m_{1}}$ to
$R^{n}$ within the iterative process. Instead, it uses the transposed
matrix $A_{1}^{T}$ which is readily available. The $CQ$-algorithm
\cite[Algorithm 1.1]{Byrne2002} is in fact a projected Landweber method for
the split feasibility formulation of Problem \ref{prob:imrt+dvc}.

In the sequel $P_{\Omega}(z)$ denotes an orthogonal projection of a vector $z$
onto a set $\Omega.$ All data quantities mentioned below are as in Problem
\ref{prob:imrt+dvc}. Since $Q$ is not a convex set there might be more than one point for $P_{Q}$ in (\ref{eq:alg:LandweberGen:1}) below, therefore, the symbol $\in$ therein means that $x^{k+1}$ could be any projection point onto $Q$ of the vector in the parentheses whose projection onto $Q$ is sought after, and can be
arbitrarily chosen from those if more then one exists.

\begin{algorithm}
\label{alg:LandweberGen}\textbf{The }$CQ$\textbf{-Algorithm for the Split
Linear Feasibility Problem with a DVC.} $\left.  {}\right.  $

\textbf{Step 0}: Take an arbitrary $x^{0}\in R^{n}$, and set $k=0$.

\textbf{Step 1}: For a current iterate $x^{k}$ calculate $A_{1}x^{k}$ and
compute the next iterate by%
\begin{equation}
x^{k+1}\in P_{C}\left(  x^{k}+\gamma A_{1}^{T}\left(  P_{Q}(A_{1}x^{k}%
)-A_{1}x^{k}\right)  \right)  . \label{eq:alg:LandweberGen:1}%
\end{equation}

If a stopping criterion applies then stop, otherwise go back to the
\textbf{Step 1} with $k\Longleftarrow k+1$.
\end{algorithm}

Next we explain how to do the projections onto $C$ and onto $Q,$ and how to choose the parameter $\gamma$ in (\ref{eq:alg:LandweberGen:1}). Since $Q$ of (\ref{eq:Q}) is not convex, the projection $P_{Q}$ may by
multivalued. Nevertheless, for any $z\in R^{m_{1}},$ we can calculate
$P_{Q}(z)$ by using the following formula%

\begin{equation}
P_{Q}(z)=P_{\overline{Q}}(z-b^{1})+b^{1} \label{eq:21}%
\end{equation}
where
\begin{equation}
\overline{Q}:=\{y\in R^{m_{1}}\mid\Vert y_{+}\Vert_{0}\leq\alpha m_{1}\}.
\label{eq:22}%
\end{equation}
Hence the projection of a point $z\in R^{m_{1}}$ onto the set $Q$ of
(\ref{eq:Q}) is obtained by projecting the shifted point $(z-b^{1})$ onto the
set $\overline{Q}$ and adding $b^{1}$ to the result. The proof of this fact can be
found in the Appendix.

Therefore, the problem reduces to computing a projection onto $\overline{Q}$. This is done according to the following recipe: First count how many
components of $(z-b^{1})$
are positive, say $\ell$. Then,%
\begin{equation}
P_{\overline{Q}}(z-b^{1})=\left\{
\begin{array}
[c]{ll}%
(z-b^{1}), & \text{If }\ell\leq\alpha m_{1},\\
w, & \text{If }\ell>\alpha m_{1},
\end{array}
\right.  \label{eq:proj-Q-bar}%
\end{equation}
where $w$ is the vector obtained from $(z-b^{1})$ by replacing its
$\ell-\alpha m_{1}$ smallest positive components by zeros and leaving the
others unchanged. If $\ell\leq\alpha m_{1}$ then the point $(z-b^{1})$ is
already inside $\overline{Q},$ thus $P_{\overline{Q}}(z-b^{1})=(z-b^{1}).$ We will use
the above for $z=$ $A_{1}x^{k}$ in
(\ref{eq:alg:LandweberGen:1}).

Following the seminal $CQ$-algorithm \cite{Byrne2002}, designed for the
case when both sets $C$ and $Q$ are convex, we propose that the parameter
$\gamma$ in (\ref{eq:alg:LandweberGen:1}) will be user-chosen from the open
interval $0<\gamma<2/\theta$ where $\theta$ is pre-calculated once. To do so we employ \cite[Corollary 2.3]{Byrne2009} by using the
squared Frobenius matrix norm $\Vert A_{1}\Vert_{F}^{2}$ and defining
\begin{equation}
\theta:=\Vert A_{1}\Vert_{F}^{2}=\sum_{i=1}^{m_{1}}%
{\displaystyle\sum\limits_{j=1}^{n}}
\left\vert a_{ij}\right\vert ^{2}, \label{eq:teta-hat}%
\end{equation}
where for $i=1,2,\ldots,m_{1}$ and $j=1,2,\ldots,n$, the entries of $A_{1}$
are $a_{ij}.$

In the practical implementation we replace the projection onto $C$
(\ref{eq:alg:LandweberGen:1}) by a sequence of projections onto the individual
inequalities of the constraints (\ref{eq:dvc1})--(\ref{eq:dvc3}) that are
collectively described by $c\leq Ax\leq b$ with where $A$, $b$ and $c$ are as
in (\ref{eq:A}), (\ref{eq:b}) and (\ref{eq:c}), respectively, according to a
feasibility-seeking projection method of our choice. All of the above leads to our proposed Dose-Volume Split-feasibility (DVSF) Algorithm.

\begin{algorithm}
\label{alg:LandweberCyclic}\textbf{The Dose-Volume Split-feasibility (DVSF) Algorithm}.$\left.  {}\right.  $\newline

\textbf{Step (-1)}: Read all data from Problem \ref{prob:imrt+dvc} and
calculate (once) the transposed matrix $A_{1}^{T},$ the value of $\theta$ according to (\ref{eq:teta-hat}), and choose a parameter $\gamma$ in the open interval
$0<\gamma<2/\theta$.

\textbf{Step 0}: Take an arbitrary $x^{0}\in R_{+}^{n}$, and set $k:=0$.

\textbf{Step 1}: Project $A_{1}x^{k}$ onto $Q$ as follows:

\textbf{Step 1.1}: For the current iterate $x^{k}$ compute $A_{1}x^{k}$, count the coordinates of $(A_{1}x^{k}-b^{1})$ that are positive and denote
their number by $\ell.$

\textbf{Step 1.2}: Calculate (using (\ref{eq:proj-Q-bar}) with $z=$
$A_{1}x^{k}$)%
\begin{equation}
v^{k}:=P_{\overline{Q}}(A_{1}x^{k}-b^{1}).
\end{equation}

\textbf{Step 1.3}: Calculate a projection of $A_{1}x^{k}$ onto $Q$ (following
(\ref{eq:21})--(\ref{eq:22})):%
\begin{equation}
P_{Q}(A_{1}x^{k})=v^{k}+b^{1}.
\end{equation}

\textbf{Step 2}: Calculate $u^{k}\in R^{n}$ by the formula
\begin{equation}
u^{k}=x^{k}+\gamma A_{1}^{T}\left(  P_{Q}(A_{1}x^{k})-A_{1}x^{k}\right)  .
\label{eq:Cstep}
\end{equation}

\textbf{Step 3}: Instead of projecting $u^{k}$ onto $C$ as required in (\ref{eq:alg:LandweberGen:1}), use $u^{k}$ from \textbf{Step 2} as an initial point and perform a sweep (or several sweeps) of a
feasibility-seeking projection method for the inequalities of (\ref{eq:dvc1}%
)--(\ref{eq:dvc4}). When stopping this sweep (or several sweeps) take the
resulting vector as the next iterate $x^{k+1}$.

\textbf{Step 4}: If a stopping criterion applies then stop, otherwise go back
to the \textbf{Step 1} with $k\Longleftarrow k+1$.
\end{algorithm}

Algorithm \ref{alg:LandweberCyclic} is a general scheme that is made specific by choosing a feasibility-seeking projection
method to be used in its \textbf{Step 3}. Consult Bauschke and Borwein \cite{bb96} for a review of such algorithms, see Censor and Cegielski \cite{cc15} for an annotated bibliography of books and reviews on the subject and Censor \textit{et al.} \cite{coap} for a review with experimental results.

We adopted here the automatic relaxation method (ARM) for feasibility-seeking \cite{arm}. We give
a generic description of this algorithm by considering the problem of solving
iteratively large and possibly sparse systems of interval linear inequalities
of the form
\begin{equation}
w_{j}\;\leq\;\left\langle a^{j},x\right\rangle \leq v_{j},\;\;\;j=1,2,...,p,
\label{eqn:slabs}%
\end{equation}
where $a^{j}\in R^{n}$ are given, for all $j$, and $w=(w_{j})\in R^{p}$, and
$v=(v_{j})\in R^{p}$ are given too. Assuming that the system is
\textit{feasible}, an $x^{\ast}\in R^{n}$ which solves (\ref{eqn:slabs}) is
required. Geometrically, the system represents $p$ nonempty hyperslabs in
$R^{n}$, each being the nonempty intersection of a pair of half-spaces. If we
are willing to ignore the slabs structure of the problem it could be addressed
as a system of $2p$ linear one-sided inequalities and solved by the Agmon-Motzkin-Schoenberg (AMS)
algorithm \cite{agmon,mot-schoen54}. The ARM takes advantage of the interval structure of the problem by
handling in every iterative step a pair of inequalities and it also
realizes a specific relaxation principle (see \cite{arm} for details) in an
automatic manner. External relaxation parameters are available on top of the
built-in automatic relaxation principle.

For every hyperslab of the system (\ref{eqn:slabs}) denote by%
\begin{equation}
\overline{H}_{j}:=\{x\in R^{n}\;|\;\left\langle a^{j},x\right\rangle
=v_{j}\}\text{ and\ }\underline{H}_{j}:=\{x\in R^{n}\;|\;\langle
a^{j},x\rangle=w_{j}\}
\end{equation}
its bounding hyperplanes. The median hyperplane will be%
\begin{equation}
H_{j}:=\{x\in R^{n}\;|\;\langle a^{j},x\rangle=\frac{1}{2}(v_{j}+w_{j})\},
\end{equation}
and the half-width $\psi_{j}$ of the hyperslab is%
\begin{equation}
\psi_{j}=\frac{v_{j}-w_{j}}{2\parallel a^{j}\parallel},
\end{equation}
where $\parallel . \parallel$ stands for the Euclidean 2-norm.
The signed distance of a point $z\in R^{n}$ from the $j$-th median hyperplane
$H_{j}$ is given by%
\begin{equation}
d(z,H_{j})=\frac{\displaystyle\langle a^{j},z\rangle-\frac{1}{2}(v_{j}+w_{j}%
)}{\displaystyle\parallel a^{j}\parallel}.
\end{equation}

Denoting $d_{j(k)}:=d(x^{k},H_{j(k)}),$ the automatic relaxation method is as follows.

\begin{algorithm}
\textbf{The} \textbf{Automatic Relaxation Method (ARM)}{.\label{algo:ARM}}

\textbf{Initialization}: $x^{0}\in R^{n}$ is arbitrary.

\textbf{Iterative step}: Given a current iterate $x^{k}$ calculate the next
iterate $x^{k+1}$ by
\begin{equation}
x^{k+1}=\left\{
\begin{array}
[c]{ll}%
x^{k}, & \text{if}\;\;|d_{j(k)}|\;\leq\;\psi_{j(k)},\\
x^{k}-\frac{\displaystyle\lambda_{k}}{\displaystyle2}\left(  \frac
{\displaystyle d_{j(k)}^{2}-\psi_{j(k)}^{2}}{\displaystyle d_{j(k)}}\right)
\frac{\displaystyle a^{j(k)}}{\displaystyle\parallel a^{j(k)}\parallel}, &
\text{otherwise.}%
\end{array}
\right.  \label{eqn:ARM}%
\end{equation}

\textbf{Control}: The control sequence $\{j(k)\}_{k=0}^{\infty}$ according to
which hyperslabs are picked during iterations is cyclic on $\{1,2,....,m\}$, i.e., $j(k) = k \bmod m+1$.

\textbf{Relaxation parameters}: External relaxation parameters $\{\lambda
_{k}\}_{k=0}^{\infty}$, available on top of the built-in automatic relaxation
principle are confined, for all $k\;\geq\;0,$ to%
\begin{equation}
\epsilon_{1}\leq\lambda_{k}\leq2-\epsilon_{2},\;\text{for some user-chosen arbitrarily small}%
\;\epsilon_{1},\epsilon_{2}>0.
\end{equation}

\end{algorithm}

\subsection{Performance Testing}
Performance tests with two different geometries were carried out to verify the functionality of the proposed algorithmic structure for IMRT. Applications to IMPT are presented in the current work. However, the algorithm is not proton specific, and is equally applicable to any form of IMRT. Only the values of the matrix $A$ differ when different forms of radiation are used.

\subsubsection{Simplified 2D C-shaped geometry}
\begin{figure}[ht]
	\begin{center}
		\includegraphics[width=1.0\textwidth]{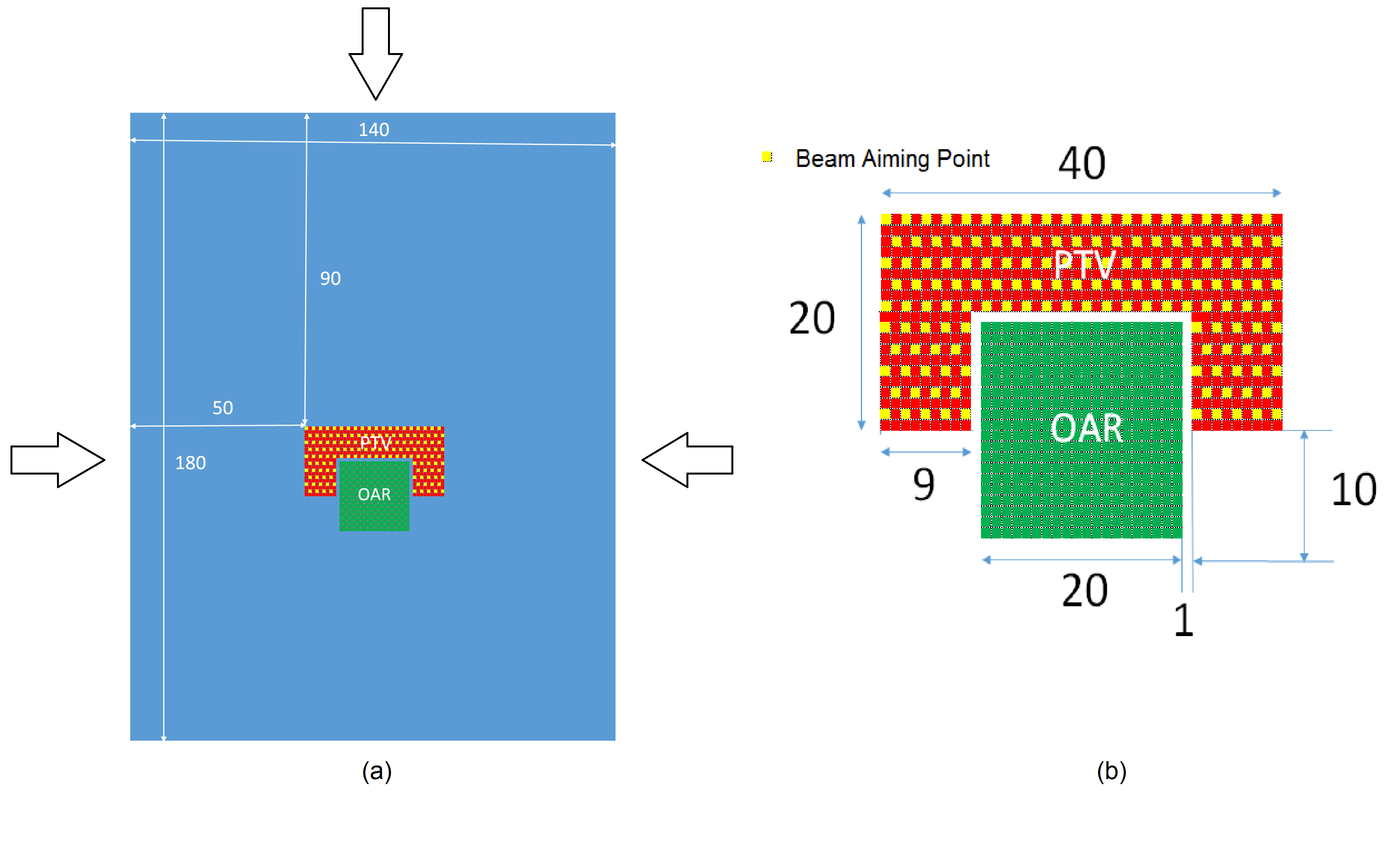}
		\caption{(a) Simplified 2D geometry simulating a tumour surrounding the brainstem. Arrows indicate the proton beam directions selected. (b) Magnified view of the target structure (PTV) and the avoidance structure (OAR). Dimensions are in millimetres. The yellow squares represent locations of delivered Bragg peaks from each beam.}
		\label{fig:geom}
	\end{center}
\end{figure}

A 2D test geometry was defined to simulate an axial cross-section of a tumour volume surrounding an organ at risk. The test geometry is illustrated in Figure \ref{fig:geom}. Structure pixels were defined with a resolution of 1 mm, also coinciding with the dose grid.

A proton pencil beamlet spacing of 2 mm, evenly distributed throughout the PTV structure, was used. Three beam angles were used to deliver dose to the PTV area. Each beam contained 146 proton pencil beamlets. The dose deposited by each pencil beamlet in the dose grid was calculated with the Monte Carlo toolkit Geant4 \cite{G4} and recorded in a text file.

The simulated beamlets were uniform circular proton beams of 2~mm
diameter. A pre-absorber made of 5.5~cm of polyethylene was inserted in the
beams at 5~cm in front of the irradiated geometry in order to smooth the Bragg
peaks and avoid dose distribution ripples due to beamlet spacing. The
beamlet energies for each aiming point were extracted from a calibration curve.
The energy used ranged from 118.5~MeV to 153~MeV with a resolution of 0.5~MeV.
The material of all the structures of the irradiated geometry was assumed to be
water.

The standard electromagnetic physics (G4EmStandardPhysics) and hadron physics
models (G4HadronPhysicsQGSP$\_$BIC$\_$HP) were used for proton tracking. Hadron
elastic scattering physics, stopping physics, ion physics and decays models were
also activated. A range cut of 0.1~mm was set for all particles. For each beamlet, $10^6$ events were simulated and the mean absorbed dose per proton was calculated at each pixel of the dose grid.

A series of dose-volume constraints (DVCs) were defined to verify the functionality of the algorithm. These included:

\begin{itemize}
	\item dose only constraints (DOCs) applied to both the PTV and OAR structures
	\item a single DVC associated with a single structure (the OAR structure)
	\item multiple (two) DVCs associated with a single structure (the OAR structure)
	\item DVCs associated with multiple structures (the PTV structure and the OAR structure)
\end{itemize}

At this point it is instructive to reconcile the dose-volume terminology used in the current work and the terminology commonly used in the literature. Let us consider an example where a prescription has been made to an OAR such that only 20\% of the volume can receive more than 40 Gy and none of the volume can receive more than 50 Gy. Using the terminology of the current work, this would correspond to values of $\alpha = 0.2$, $b_1$ = 40 Gy, and $\beta = 0.25$ in Problem \ref{prob:imrt+dvc}. Using the common terminology, this would correspond to $D_{20\%} \leq 40$ Gy and $D_{max} = 50$ Gy.

Table \ref{tab:prescriptions} lists the combination of DVCs enforced in the current work, using the common terminology of IMRT DVCs. The dose prescriptions and percentage volume violations were chosen to allow for a demonstration of the functionality of the algorithm.

The initial pencil beam intensity vector before inverse planning was set to unity. The dose distribution resulting from the initial intensity vector is shown in Figure \ref{fig:dose_maps}(a). The proposed algorithm was run for 2000 cycles for each prescription listed in Table \ref{tab:prescriptions}. In this terminology, one cycle corresponds to one complete processing of all DVCs and DOCs applied to each pixel within both the PTV and OAR structures.

\begin{table}[ht]
	\caption{Prescriptions associated with PTV and OAR structures in order to test the functionality of the proposed DVSF algorithm (Algorithm \ref{alg:LandweberCyclic}) in a simplified 2D geometry.}
	\begin{center}
		\begin{tabular}{ccc}
			\hline
			Prescription & OAR & PTV \\
			\hline
			\\
			1 & ${D_{max} = 45}$ Gy & \parbox{3cm}{${D_{min} = 70}$ Gy \\ ${D_{max} = 77}$ Gy} \\
			\\
			\hline
			\\
			2 & \parbox{3cm}{${D_{15.5\%} \leq 45}$ Gy \\ ${D_{max} = 70}$ Gy} & \parbox{3cm}{${D_{min} = 70}$ Gy \\ ${D_{max} = 77}$ Gy} \\
			\\
			\hline
			\\
			3 & \parbox{3cm}{${D_{50\%} \leq 25}$ Gy \\ ${D_{15.5\%} \leq 45}$ Gy \\ ${D_{max} = 70}$ Gy} & \parbox{3cm}{${D_{min} = 70}$ Gy \\ ${D_{max} = 77}$ Gy} \\
			\\
			\hline
			\\
			4 & \parbox{3cm}{${D_{8.5\%} \leq 45}$ Gy \\ ${D_{max} = 70}$ Gy} & \parbox{3cm}{${D_{min} = 66.5}$ Gy \\ ${D_{95\%} \geq 70}$ Gy \\ ${D_{max} = 77}$ Gy} \\
			\\
			\hline
		\end{tabular}
	\end{center}
	\label{tab:prescriptions}
\end{table}

\subsubsection{Clinical 3D geometry}
In keeping with the 2D geometry, a base of skull chordoma IMPT treatment plan was chosen due to the challenging constraints imposed by a target structure surrounding an avoidance structure. The \emph{Philips} \emph{Pinnacle}$^3$ treatment planning system (Philips Healthcare, Koninklijke Philips N.V.) was used to contour the PTV and brainstem. The exported DICOM RT (structure) files were imported into a MATLAB (The MathWorks, Inc.) script and the brainstem and PTV contours were mapped over the CT coordinates. A dose grid was created in MATLAB to match that defined in \emph{Pinnacle}$^3$. The dimensions were $42 \times 43 \times 9$ voxels with resolutions of 2 mm, 2 mm and 3 mm in the $x$, $y$ and $z$ dimensions, respectively. The dose grid was twice as large as the CT pixel size in the $x$ and $y$ dimension and equivalent to the CT resolution in the $z$ dimension. A reduced number of slices (9) was required due to memory restrictions encountered during the export of pencil beamlet doses.

An IMPT treatment plan was created in the \emph{Pinnacle}$^3$ research release of proton pencil beam scanning (PBS). Two beams were targeted at the PTV from angles of 80$^o$ and 280$^o$, containing 574 and 564 beamlets, respectively. A range shifter of 7.5 cm thickness was used with both beams to ensure proximal PTV coverage. Distal and proximal margins for pencil beam placement were automatically calculated as a percentage of proton range. The dose grid resulting from each unit intensity beamlet was exported from \emph{Pinnacle}$^3$. Beamlet parameters were set to 80$\%$ layer overlap, a lateral spot resolution of 0.6 cm, a lateral target margin of 0.4 cm and 3 standard deviation dose spread during dose calculation. Dose was calculated with the analytical PBS algorithm which includes nuclear attenuation and an energy and material dependent multiple Coulomb scattering model.


For each structure $A$-matrices were created by combining the geometry defined by the DICOM RT structures and the dose grid obtained for each beamlet. Each 3D beamlet dose grid was rearranged to a 1D vector which became a column of an $A$-matrix. Each row of the $A_{OAR}$ matrix corresponded to a voxel of the brainstem and likewise each row of the $A_{PTV}$ matrix corresponded to a voxel of the PTV.

Two DVCs were tested for the base of skull chordoma IMPT treatment plan (see Table \ref{tab:prescriptionsClinical}). The DVCs differed in the dose objectives for the brainstem while keeping the PTV objectives constant. The same DVCs were applied consistently for both the DVSF algorithm (Algorithm \ref{alg:LandweberCyclic}) and \emph{Pinnacle}$^3$.

\begin{table}[ht]
	\caption{Prescriptions associated with PTV and OAR (brainstem) structure for a clinical test case.}
	\begin{center}
		\begin{tabular}{c | c c}
			\hline
			Prescription & OAR & PTV \\
			\hline
			\\
			1 & $D_\text{max} = 54$ Gy & $D_\text{min} = 66.5$ Gy \\
			& $D_{5\%} \leq 50$ Gy & $D_\text{max} = 74.9$ Gy \\
			& & $D_{95\%} \geq 70$ Gy \\
			\\
			\hline \\
			2 & $D_\text{max} = 40$ Gy & $D_\text{min} = 66.5$ Gy \\
			& $D_{5\%} \leq 35$ Gy & $D_\text{max} = 74.9$ Gy \\
			& & $D_{95\%} \geq 70$ Gy \\
			 \\
			\hline
		\end{tabular}\\
		\label{tab:prescriptionsClinical}
	\end{center}
\end{table}

Independent values for the parameter $\gamma$ of (\ref{eq:Cstep}) were used for the OAR and PTV and are denoted by $\gamma_{PTV}$ and $\gamma_{OAR}$. These values were determined from the structure-specific calculation of $\theta$ in (\ref{eq:teta-hat}), denoted by $\theta_{PTV}$ and $\theta_{OAR}$. The relaxation parameters $\lambda_k$ of (\ref{eqn:ARM}) are fixed throughout the iterations and represented by $\lambda_{PTV}$ and $\lambda_{OAR}$.

\section{Results}
\subsection{Simplified 2D C-shaped geometry}
\begin{figure}[htp]
	\begin{center}
		\includegraphics[width=1.0\textwidth]{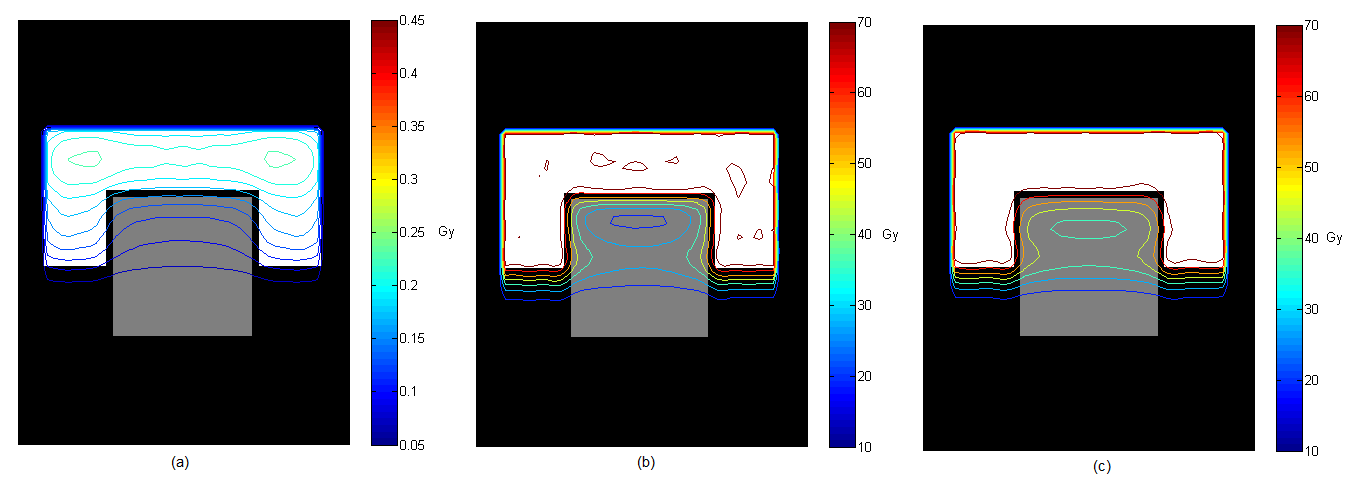}
		\caption{Isodose contours corresponding to (a) unit intensity pencil beams, (b) dose only constraints (Prescription 1), and (c) DVCs applied to both the PTV structure and the OAR structure (Prescription 4).}
		\label{fig:dose_maps}
	\end{center}
\end{figure}

\begin{figure}[htp]
	\begin{center}
		\includegraphics[width=1.0\textwidth]{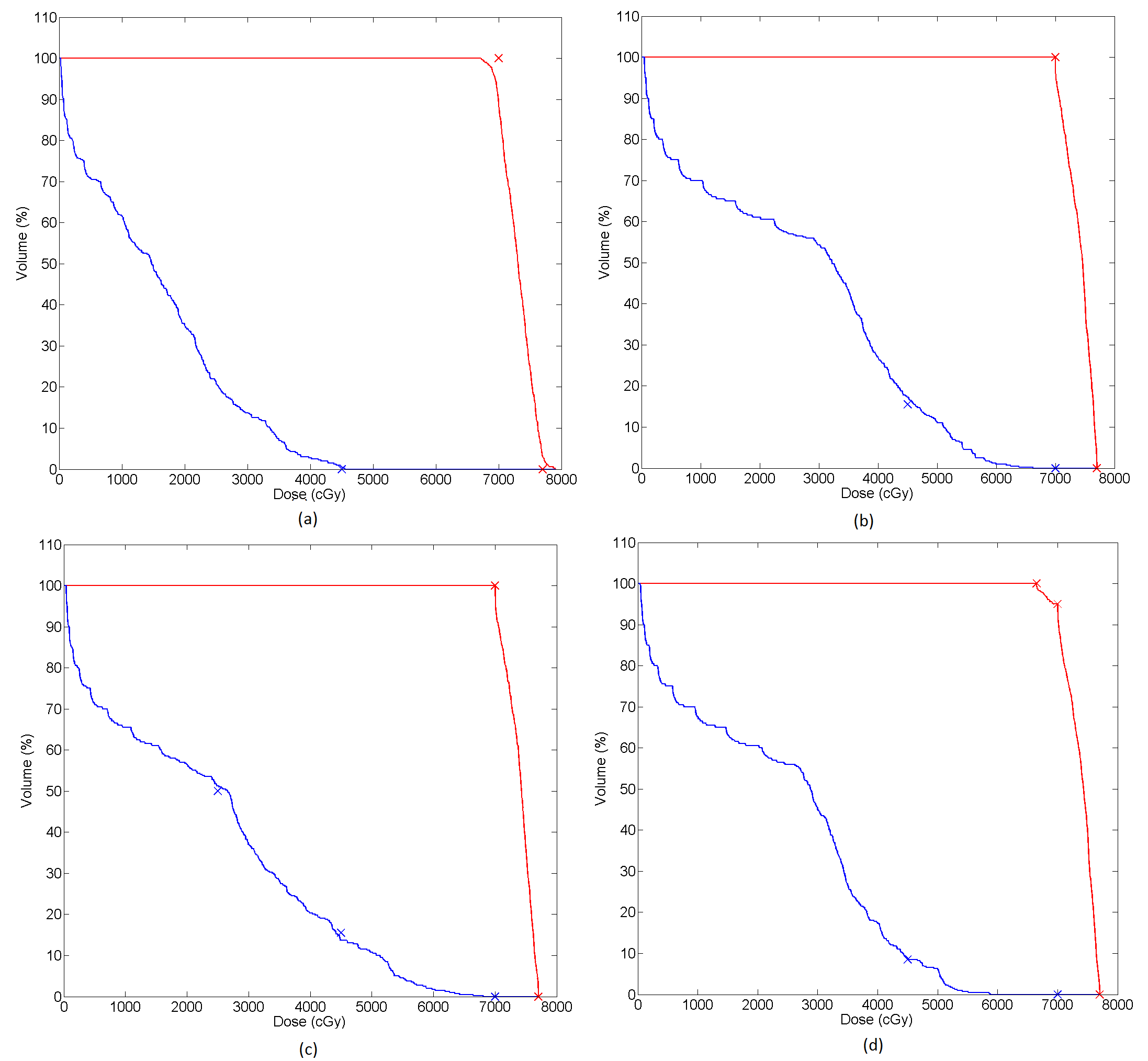}
		\caption{Dose-volume histograms for each of the prescriptions tested in the 2D C-shaped geometry. (a) Prescription 1 (b) Prescription 2 (c) Prescription 3 (d) Prescription 4. Red lines correspond to the PTV and blue lines to the OAR. Crosses indicate the DVCs.}
		\label{fig:DVHs}
	\end{center}
\end{figure}

The dose distributions following inverse planning for Prescriptions 1 and 4 in Table \ref{tab:prescriptions} are shown in Figure \ref{fig:dose_maps}(b) and \ref{fig:dose_maps}(c). The dose-volume histograms following inverse planning for all cases listed in Table \ref{tab:prescriptions} are presented in Figure \ref{fig:DVHs}.

The dose distributions (Figure \ref{fig:dose_maps}) allow for a qualitative assessment of the functionality of the DVSF algorithm (Algorithm \ref{alg:LandweberCyclic}). It is evident that the dose resulting from unit intensity pencil beamlets is successfully modulated toward the desired dose distribution. However, for a quantitative assessment the dose-volume histograms must be considered. When Prescription 1 DOCs were applied the dose objectives on the PTV structure could not be met (Figure \ref{fig:DVHs}(a)). Introducing the DVC on the OAR structure relaxed these conditions and resulted in satisfaction of the dose objectives on the PTV structure (Figure \ref{fig:DVHs}(b)). While the DVC on the OAR structure was not achieved in Prescription 2, continued iterations would have resulted in a dose distribution approaching the DVC more closely. The DVSF algorithm (Algorithm \ref{alg:LandweberCyclic}) was shown to function with multiple DVCs applied to a single structure (Figure \ref{fig:DVHs}(c)), and with DVCs applied to multiple structures (Figure \ref{fig:DVHs}(d)).

\subsection{Clinical 3D geometry}
\begin{figure}[htp]
	\begin{center}
		\includegraphics[width=1.0\textwidth]{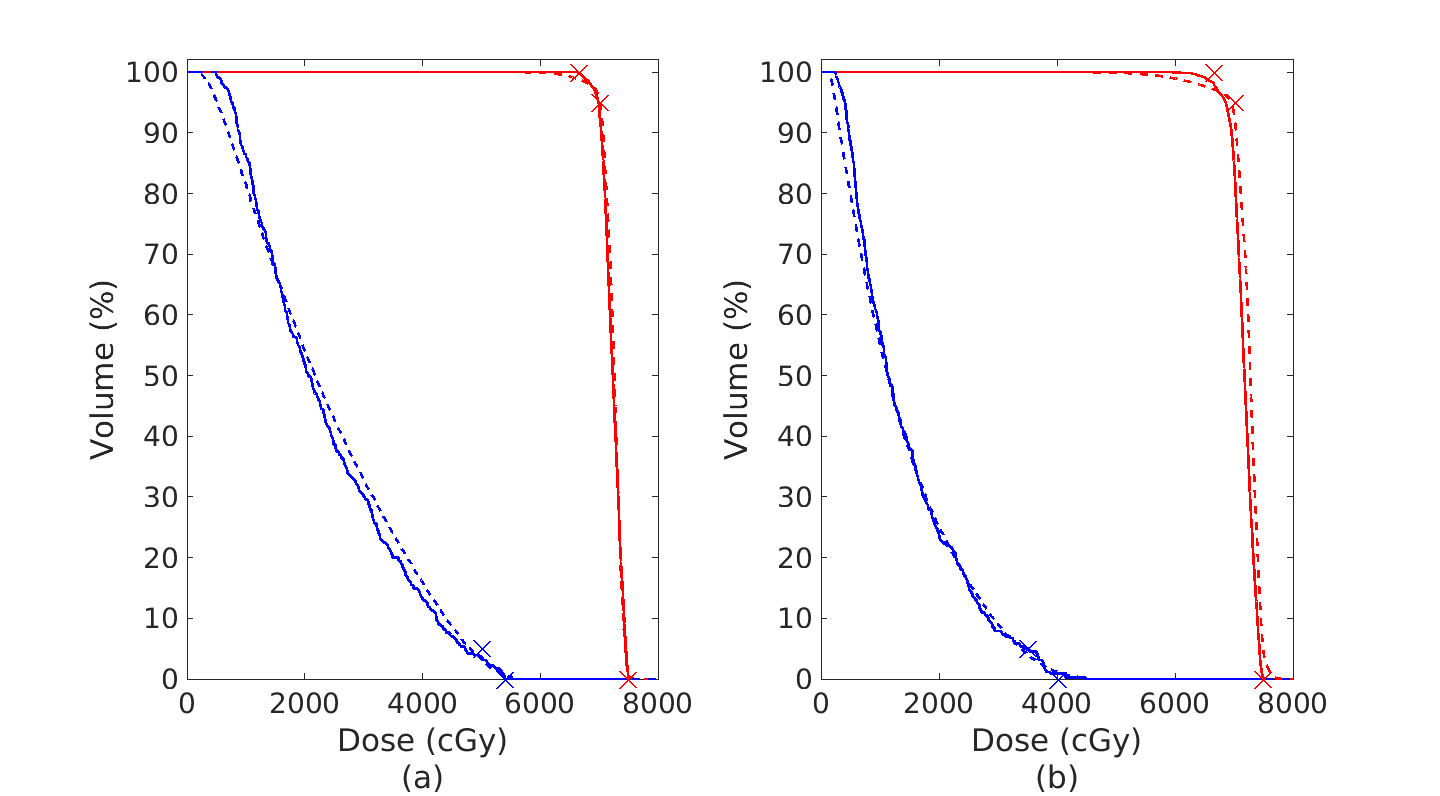}\\
		\caption{(a) Prescription 1 of Table \ref{tab:prescriptionsClinical}: DVH after 2000 cycles of the DVSF algorithm (Algorithm \ref{alg:LandweberCyclic}) (solid), using $\gamma_\text{PTV} = 1.99/\theta_\text{PTV}$, $\gamma_\text{OAR} = 1/\theta_\text{OAR}$, $\lambda_\text{PTV} =\lambda_\text{OAR} = 1$, compared to the DVH produced by \emph{Pinnacle}$^3$ (dashed) after 86 iterations and meeting a stopping tolerance of less than $10^{-7}$. (b) Prescription 2 of Table \ref{tab:prescriptionsClinical}: DVH after 2000 cycles of the DVSF algorithm (Algorithm \ref{alg:LandweberCyclic}) (solid), using $\gamma_\text{PTV} = 1.99/\theta_\text{PTV}$, $\gamma_\text{OAR} = 0.3/\theta_\text{OAR}$, $\lambda_\text{PTV} =\lambda_\text{OAR} = 0.5$, compared to the DVH produced by \emph{Pinnacle}$^3$ (dashed) after 131 iterations and meeting a stopping tolerance of less than $10^{-7}$.}
		\label{fig:ClinicalDVH1}
	\end{center}
\end{figure}

\begin{figure}[htp]
\begin{center}
	\includegraphics[height=0.8\textheight]{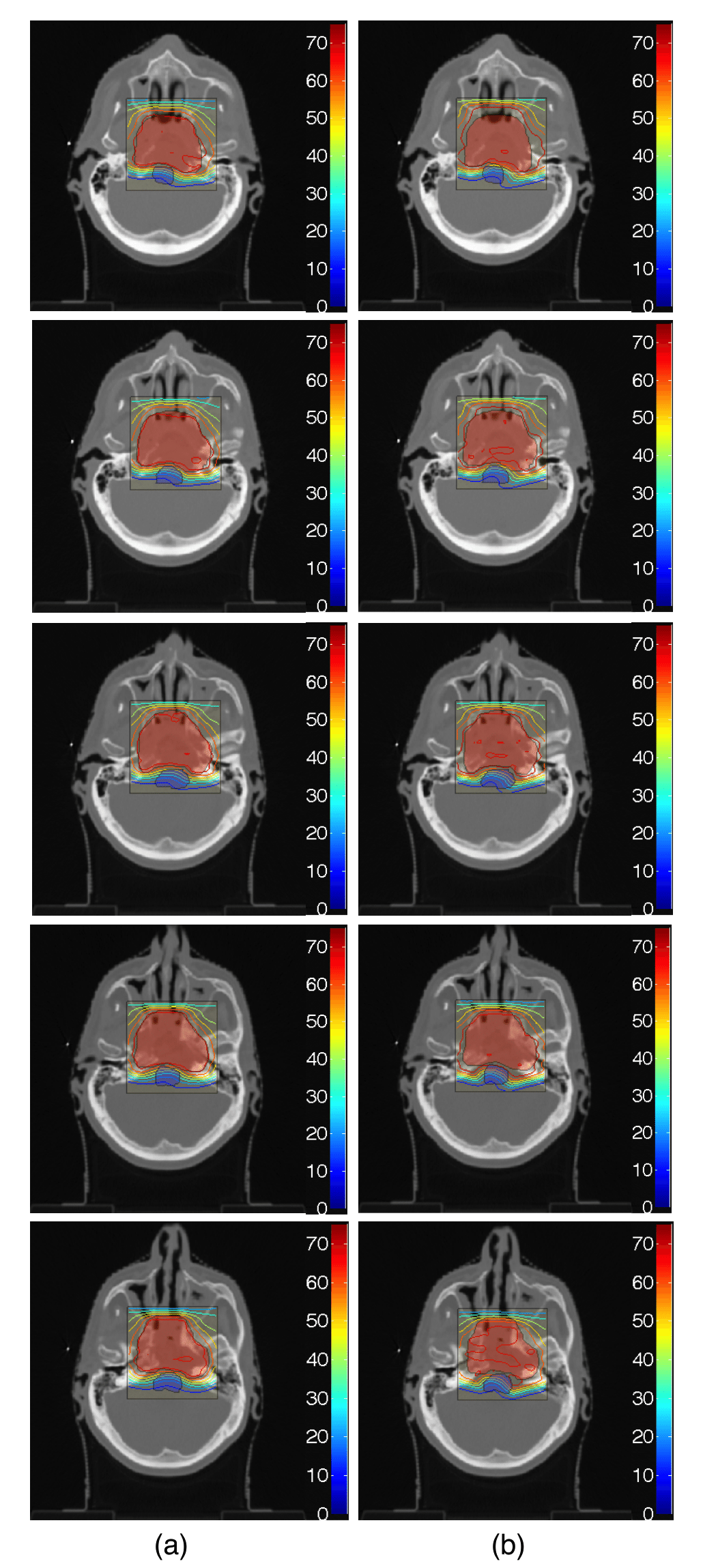}\\
	\caption{Dose contour map for Prescription 1 of Table \ref{tab:prescriptionsClinical} (a) after 2000 cycles of the DVSF algorithm (Algorithm \ref{alg:LandweberCyclic}), using $\gamma_\text{PTV} = 1.99/\theta_\text{PTV}$, $\gamma_\text{OAR} = 1/\theta_\text{OAR}$, $\lambda_\text{PTV} =\lambda_\text{OAR} = 1$, compared to (b) the dose contour map produced by \emph{Pinnacle}$^3$ after 86 iterations and meeting a stopping tolerance of less than $10^{-7}$. The red shaded area is the PTV and the blue shaded area is the OAR (brainstem). Colour bar has units of Gy.}
	\label{fig:ClinicalMap1}
\end{center}
\end{figure}

Cumulative DVHs for Prescription 1 of Table \ref{tab:prescriptionsClinical} using the DVSF algorithm (Algorithm \ref{alg:LandweberCyclic}) and that produced by the \emph{Pinnacle}$^3$ inverse planning algorithm are shown in Figure \ref{fig:ClinicalDVH1}. All constraints of the less challenging dose objectives were met by the DVSF algorithm (Algorithm \ref{alg:LandweberCyclic}) whereas \emph{Pinnacle}$^3$ exceeded the maximum dose for the OAR and did not satisfy the PTV minimum dose constraint. It should be noted that the \emph{Pinnacle}$^3$ inverse planning was run only once with unit weighting on all dose objectives. It is possible that alteration of the objective weightings by trial-and-error may have resulted in a more desirable dose distribution. However, the objective of the current work was to compare the inherent ability of the algorithms to satisfy the inverse problem, and as such, iterative plan refinement by altering objective weights was not considered. Dose distributions for Prescription 1 of Table \ref{tab:prescriptionsClinical} are shown in a single axial slice in Figure \ref{fig:ClinicalMap1}. The DVSF algorithm (Algorithm \ref{alg:LandweberCyclic}) showed higher conformality of the target structure.

\begin{figure}[htp]
	\begin{center}
		\includegraphics[width=1.0\textwidth]{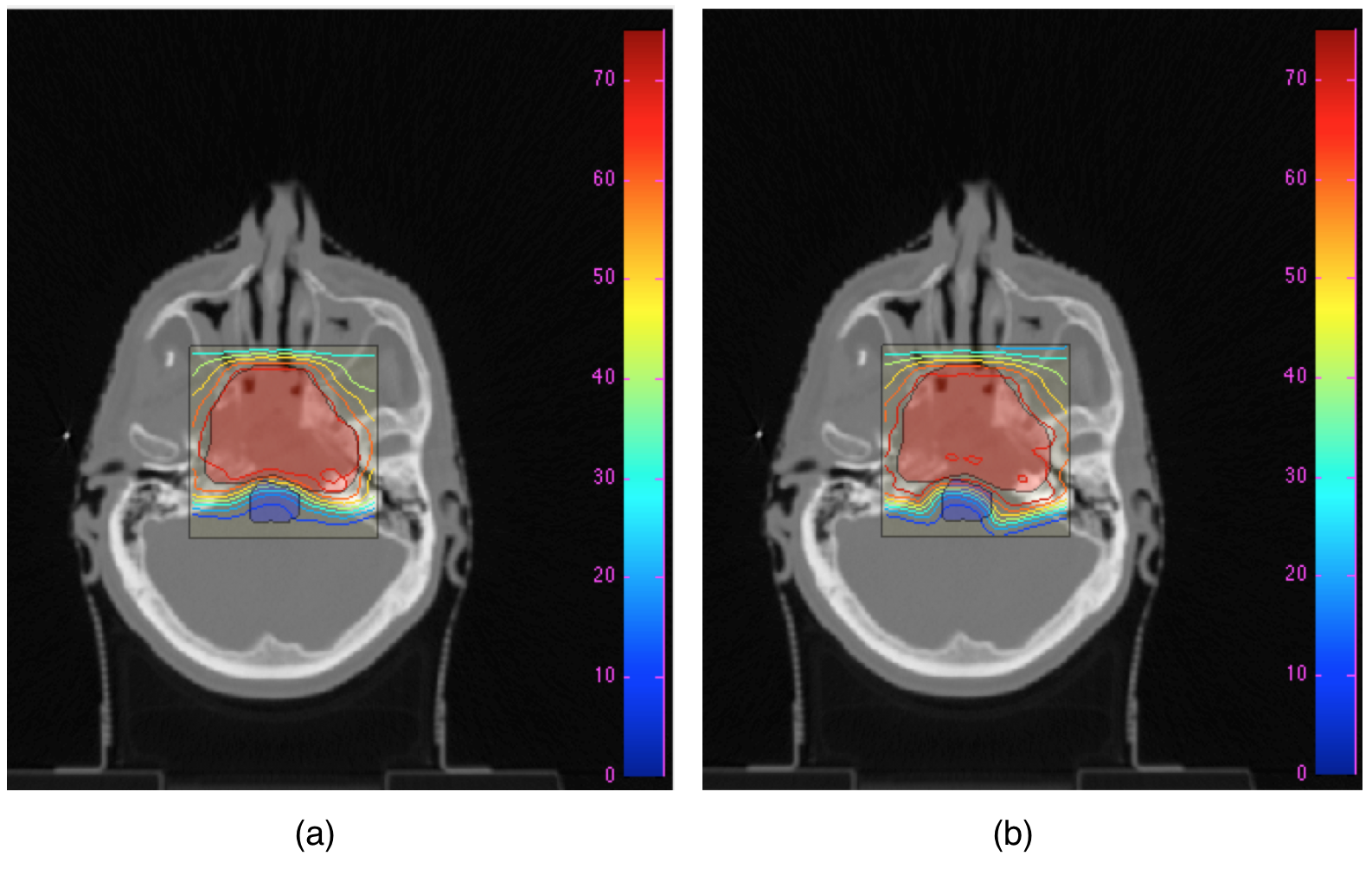}\\
		\caption{Dose contour map for Prescription 2 of Table \ref{tab:prescriptionsClinical} (a) after 2000 cycles of the DVSF algorithm (Algorithm \ref{alg:LandweberCyclic}), using $\gamma_\text{PTV} = 1.99/\theta_\text{PTV}$, $\gamma_\text{OAR} = 0.3/\theta_\text{OAR}$, $\lambda_\text{PTV} =\lambda_\text{OAR} = 0.5$, compared to (b) dose contour maps produced by \emph{Pinnacle}$^3$ after 131 iterations and meeting a stopping tolerance of less than $10^{-7}$. The red shaded area is the PTV and the blue shaded area is the OAR (brainstem). Colour bar has units of Gy.}
		\label{fig:ClinicalMap2}
	\end{center}
\end{figure}

Cumulative DVHs for Prescription 2 of Table \ref{tab:prescriptionsClinical} are shown in Figure \ref{fig:ClinicalDVH1}(b). It is clear that both the DVSF algorithm (Algorithm \ref{alg:LandweberCyclic}) and \emph{Pinnacle}$^3$ had more difficulty meeting the dose objectives in this case. The DVSF algorithm (Algorithm \ref{alg:LandweberCyclic}) was better able to meet the hard dose constraints when compared to \emph{Pinnacle}$^3$ but the latter was closer to meeting the $D_{95}\geq 70$ Gy DVC on the PTV. Dose distributions for Prescription 2 of Table \ref{tab:prescriptionsClinical} are shown in a single axial slice in Figure \ref{fig:ClinicalMap2}. Both dose distributions show cold spots in the target region.

\section{Discussion and Conclusion}
A new DVSF algorithm (Algorithm \ref{alg:LandweberCyclic}) based on feasibility-seeking has been successfully applied to IMPT inverse planning in the current work. The proposed DVSF algorithm (Algorithm \ref{alg:LandweberCyclic}) is based on a modification of the $CQ$-algorithm of Byrne \cite{Byrne2002} and is capable of directly incorporating the DVCs associated with radiation therapy prescriptions into the split feasibility-seeking problem statement. Our DVSF algorithm (Algorithm \ref{alg:LandweberCyclic}) is not restricted to IMPT and is equally applicable to other forms of IMRT inverse planning. Test cases consisted of a simplified 2D C-shape target surrounding an avoidance structure and a clinical base of skull chordoma abutting the brainstem.

The DVSF algorithm (Algorithm \ref{alg:LandweberCyclic}) performs orthogonal projections to satisfy both the DVCs and the lower and upper dose constraints. The AMS cyclic projection method \cite{agmon} was implemented for single-sided inequality dose objectives and the ARM algorithm of \cite{arm} was implemented for interval inequalities (i.e., upper and lower dose bounds for a given structure).

A series of experiments were performed with 2D C-shaped geometry using varying DVCs to validate the functionality of our DVSF algorithm (Algorithm \ref{alg:LandweberCyclic}). While DVC aims were not met in all cases within the allowed number of iterations, the shape of the DVH curve verified that the algorithm was attempting to meet these objectives. Experimentation with user-defined relaxation parameter values $\gamma$ and $\lambda$ was performed to investigate the effect of these settings on algorithmic performance. When $\lambda$ was left at the fixed value of 1, it was found that $\gamma$ values closer to the upper allowable limit of $2/\theta$ were required to meet the DVC aims. Further work concerning automatic choice of these user-defined parameters is currently being undertaken and will be presented in future investigations.

A clinical 3D IMPT treatment geometry was also investigated. The performance of the DVSF algorithm (Algorithm \ref{alg:LandweberCyclic}) was compared to that of the research release of \emph{Pinnacle}$^3$ with proton pencil beam scanning. The shape of DVHs differed for the two inverse planning algorithms. For the prescriptions investigated, our DVSF algorithm (Algorithm \ref{alg:LandweberCyclic}) was found to result in a more conformal dose distribution when assessing isodose contours and DVH distributions. It is acknowledged that the dose distributions obtained with \emph{Pinnacle}$^3$ may be improved with the addition of planning structures. However, to allow for a comparison of the inverse planning algorithms directly, no such structures were included in the treatment planning method.

While the implementation of the DVSF algorithm (Algorithm \ref{alg:LandweberCyclic}) was sequential in the current work, the structure of the algorithm lends itself to parallelization. For example, block-iterative or string-averaging projection operators may be used when performing the orthogonal projections described in Step 3 of Algorithm \ref{alg:LandweberCyclic}. Such implementations will not only have benefits in computational speed, but may also result in superior dose distributions, as has been observed in the use of these algorithms in tomographic image reconstruction \cite{scott-et-al-10}. Further work will examine the potential of block-iterative and string-averaging algorithmic schemes for the DVSF algorithm (Algorithm \ref{alg:LandweberCyclic}).

\section{Appendix}

Here is a proof of formula (\ref{eq:21}) for the projection calculation onto
the non-convex set $Q$.

\begin{proof}
We show that the following translation formula%
\begin{equation}
P_{Q}(z)=P_{\overline{Q}}(z-b^{1})+b^{1} \label{eq:21:proof:1}%
\end{equation}
holds true for every $z\in R^{m_{1}}$, despite the fact that $P_{Q}$ and
$P_{\overline{Q}}$ are set-valued, i.e., a point $z\in R^{m_{1}}$ might have more
than one projection onto the set. Note that
\begin{equation}
\overline{Q}=Q-b^{1}. \label{eq:21:proof:2}%
\end{equation}
By the definition of projection of a point onto a set,%
\begin{equation}
q_{0}\in P_{Q}(z)\text{ if and only if }q_{0}\in Q\text{ and }\Vert z-q_{0}\Vert\leq\Vert
z-q\Vert\text{, for all }q\in Q. \label{eq:21:proof:3}%
\end{equation}
Similarly, $(q_{0}-b^{1})\in P_{\overline{Q}}(z-b^{1})$ if and only if $(q_{0}-b^{1})\in\hat
{Q}$ and%
\begin{equation}
\Vert(z-b^{1})-(q_{0}-b^{1})\Vert\leq\Vert(z-b^{1})-\overline{q}\Vert
\label{eq:21:proof:4}%
\end{equation}
holds for every $\overline{q}\in\overline{Q}$. Therefore, by (\ref{eq:21:proof:2}),
(\ref{eq:21:proof:3}) and (\ref{eq:21:proof:4}), we have the following
equivalences
\begin{align}
q_{0}\in P_{Q}(z)  &  \Longleftrightarrow q_{0}\in Q\ \text{and}%
\ \forall_{q\in Q}\text{ }\Vert z-q_{0}\Vert\leq\Vert z-q\Vert\text{
}\nonumber\\
&  \Longleftrightarrow(q_{0}-b^{1})\in\overline{Q}\ \text{and}\nonumber\\
&  \forall_{q\in Q}\ \Vert(z-b^{1})-(q_{0}-b^{1})\Vert\leq\Vert(z-b^{1}%
)-(q-b^{1})\Vert\nonumber\\
&  \Longleftrightarrow(q_{0}-b^{1})\in\overline{Q}\ \text{and}\nonumber\\
&  \forall_{\overline{q}\in\overline{Q}}\ \Vert(z-b^{1})-(q_{0}-b^{1})\Vert\leq
\Vert(z-b^{1})-\overline{q}\Vert\nonumber\\
&  \Longleftrightarrow(q_{0}-b^{1})\in P_{\overline{Q}}(z-b^{1})\nonumber\\
&  \Longleftrightarrow q_{0}\in P_{\overline{Q}}(z-b^{1})+b^{1},
\end{align}
which completes the proof.
\end{proof}

\textbf{Acknowledgments}. We thank the two anonymous referees for their constructive comments which helped us improve the paper. The work of Y. Censor and R. Schulte was supported by Research Grant No. 2013003 of the United States-Israel Binational Science Foundation (BSF) and by Award No. 1P20183640-01A1 of the National Cancer Institute (NCI) of the National Institutes of Health (NIH). The authors thank \emph{Philips} for technical assistance with \emph{Pinnacle}$^3$ software for this research. 

\bibliographystyle{ieeetr}
\bibliography{IMPT-DVC}

\end{document}